\documentclass[twocolumn,aps,apl,floatfix,superscriptaddress]{revtex4-1}
\usepackage{amsmath,amssymb}
\usepackage{graphicx,float}
\usepackage{times,txfonts}
\usepackage{color}
\usepackage{multirow}
\usepackage{bbold}
\usepackage{color}
\usepackage{soul}
\usepackage{hyperref}
\usepackage{times,txfonts}
\usepackage{natbib}

\newcommand{\ket}[1]{\left| #1 \right\rangle}

\newcommand{\qo}[1]{``#1''}

\renewcommand{\epsilon}{\varepsilon}
\renewcommand{\phi}{\varphi}

\usepackage{sistyle}
\usepackage{soul}
\definecolor{lightblue}{RGB}{185,210,248}
\sethlcolor{lightblue}

\begin{document}
\title{Optical spin-to-orbital angular momentum conversion in ultra-thin metasurfaces with arbitrary topological charges}
\author{Fr\'ed\'eric Bouchard}
\affiliation{Department of Physics, University of Ottawa, 25 Templeton, Ottawa, Ontario, K1N 6N5 Canada}
\author{Israel De Leon}
\affiliation{Department of Physics, University of Ottawa, 25 Templeton, Ottawa, Ontario, K1N 6N5 Canada}
\author{Sebastian A. Schulz}
\affiliation{Department of Physics, University of Ottawa, 25 Templeton, Ottawa, Ontario, K1N 6N5 Canada}
\author{Jeremy Upham}
\affiliation{Department of Physics, University of Ottawa, 25 Templeton, Ottawa, Ontario, K1N 6N5 Canada}
\author{Ebrahim Karimi}
\email{ekarimi@uottawa.ca}
\affiliation{Department of Physics, University of Ottawa, 25 Templeton, Ottawa, Ontario, K1N 6N5 Canada}
\author{Robert W Boyd}
\affiliation{Department of Physics, University of Ottawa, 25 Templeton, Ottawa, Ontario, K1N 6N5 Canada}
\affiliation{Institute of Optics, University of Rochester, Rochester, New York, 14627, USA}

\begin{abstract}
Orbital angular momentum associated with the helical phase-front of optical beams provides an unbounded \qo{space} for both classical and quantum communications. Among the different approaches to generate and manipulate orbital angular momentum states of light, coupling between spin and orbital angular momentum allows a faster manipulation of orbital angular momentum states because it depends on manipulating the polarisation state of light, which is simpler and generally faster than manipulating conventional orbital angular momentum generators. In this work, we design and fabricate an ultra-thin spin-to-orbital angular momentum converter, based on plasmonic nano-antennas and operating in the visible wavelength range that is capable of converting spin to an arbitrary value of OAM $\ell$. The nano-antennas are arranged in an array with a well-defined geometry in the transverse plane of the beam, possessing a specific integer or half-integer topological charge $q$. When a circularly polarised light beam traverses this metasurface, the output beam polarisation switches handedness and the OAM changes in value by $\ell = \pm2q\hbar$ per photon. We experimentally demonstrate $\ell$ values ranging from $\pm 1$ to $\pm 25$ with conversion efficiencies of $8.6\pm0.4~\%$. Our ultra-thin devices are integratable and thus suitable for applications in quantum communications, quantum computations and nano-scale sensing. 
\end{abstract}
\pacs{Valid PACS appear here}
\maketitle

In addition to linear momentum and energy, light beams also possess angular momentum~\cite{allen:03}. It has been shown that in vacuum the optical angular momentum can be decomposed into two nominally independent terms; spin angular momentum (SAM) and orbital angular momentum (OAM)~\cite{darwin:32,humblet:43}. SAM is intrinsic and related to the vectorial nature of light. A circularly polarised beam carries SAM: $+\hbar$ per photon for a left-circularly polarised beam and $-\hbar$ per photon for a right-circularly polarised beam, where $\hbar$ stands for the reduced Planck constant~\cite{beth:36}. By comparison, OAM has both intrinsic and extrinsic terms, the latter of which is coordinate dependent~\cite{o:02}. The intrinsic OAM, hereafter simply referred to as OAM, is related to the azimuthal dependence of the optical phase. Thus, a beam with a helical wavefront that has a phase dependence of $\exp(i\ell\phi)$ carries OAM of $\ell\hbar$ per photon, where $\ell$ can take any positive or negative integer value~\cite{allen:92}.
Although, SAM and OAM are two \qo{rotational} degrees of freedom of light that are nearly independent, they can interact under specific conditions, such as tight focusing of a circularly polarised light beam~\cite{zhao:07} and light-matter interaction in space-variant dielectric~\cite{bomzon:02}, inhomogeneous birefringent~\cite{marrucci:06} or structured media~\cite{li:13,karimi:14c}. We recently demonstrated a metasurface composed of an inhomogenous array of plasmonic nano-antennas with a thickness much smaller than the operating wavelength. This metasurface can be wielded to transform a circularly polarised light beam into a beam carrying OAM. The dimension of the individual nano-antenna  and the periodicity of the array were chosen such that the array exhibits approximately a half-wavelength, i.e. $\pi$, optical retardation between two plasmonic resonances~\cite{karimi:14c}. 
In the previous study, the plasmonic nano-antenna arrangement was cylindrically symmetric, meaning that in one cycle around the origin of the metasurface the individual nano-antennas exhibit a single rotation, which corresponds to a unit topological charge $q$ (see $q=1$ in Fig. 1 a). The device's operation principle was based on the \emph{full} spin-to-orbital angular momentum conversion. Because the metasurface was rotationally symmetric, there was \emph{no} exchange of angular momentum between the photons and the metasurface and this device yielded light carrying OAM of $\ell=\pm2q\hbar=\pm2\hbar$. However, this symmetry is not a requirement for spin to OAM conversion and a more general geometry of metasurfaces with different $q$ values could access a much broader range of OAM values.

In this work, we design and fabricate ultra-thin plasmonic metasurfaces with a general geometry possessing integer or half-integer topological charges. The working principle of these devices is based on the coupling between optical spin and OAM, an effect that occurs in inhomogeneous wave-plates~\cite{marrucci:11}. Indeed, the device flips the SAM of the input beam and the output beam gains a nonuniform, helical phase-front, depending on the orientation of the nano-antenna array. Our fabricated metasurfaces are capable of generating OAM values in the domain of $\{-25,\ldots,+25\}$ with a conversion efficiency of about $9\%$ at visible wavelengths $760~\text{nm}-790~\text{nm}$. However, the metasurface has a theoretical conversion efficiency as high as $44\%$, with reduction in efficiency originating from discrepancies between the ideal and fabricated nano-antenna dimensions. Such an efficiency would be suitable for performing recently demonstrated novel quantum computations with photon's SAM and OAM~\cite{vincenzo:13,cardano:14}. \newline

\begin{figure*}[!ht]
 \center
  \includegraphics[width=1\textwidth]{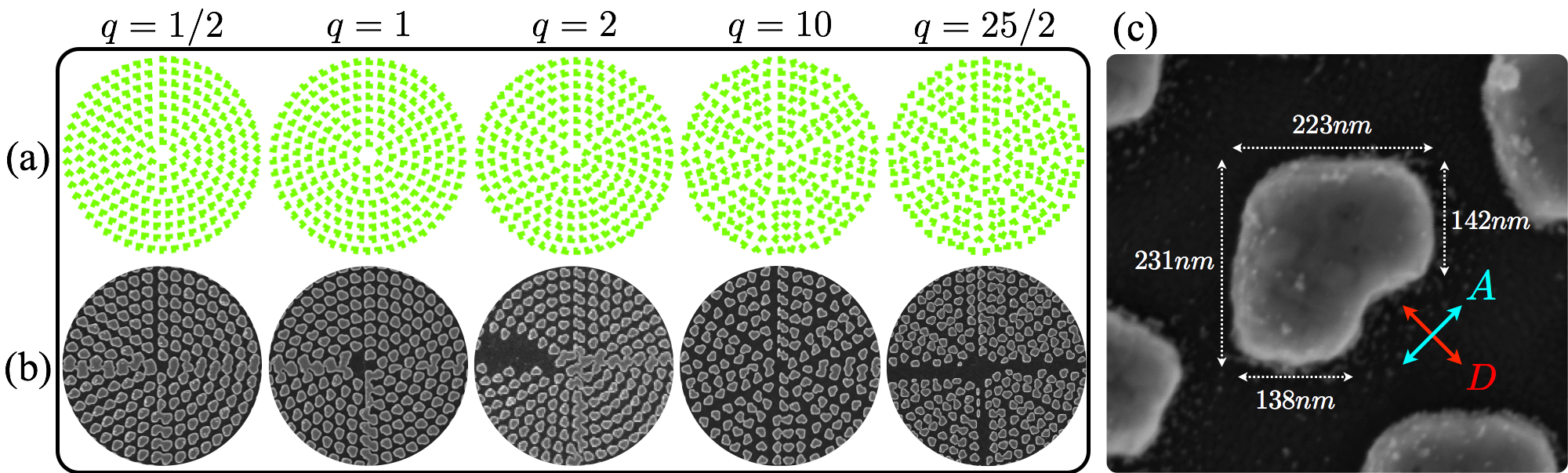}
  \caption{\label{fig:fig1} {\bf (a)} Spatial arrangement of nano-antennas in the plane of the metasurfaces for generating light beams carrying OAM values of $\{\pm1,\pm2,\pm3,\pm20$ and $\pm25\}$. {\bf (b)} Scanning electron microscope (SEM) images of the fabricated metasurfaces corresponding to the designs shown  in {\bf (a)}. $q$ indicates the topological charge. {\bf (c)} SEM image of a single L-shaped nano-antenna with a measured dimension shown in magenta colour. Diagonal ($D$) and anti-diagonal ($A$) polarisation states excite a symmetric and an anti-symmetric mode, respectively. 
  }
\end{figure*}
To generate OAM through spin-to-orbital angular momentum coupling, we require a metasurface composed of birefringent elements that exhibit a half-wavelength optical retardation between two orthogonal polarisation states. In our design, such birefringent elements take the form of L-shaped gold nano-antennas with equal arm-lengths, as illustrated in Fig.~\ref{fig:fig1}-({\bf c}). A nano-antenna with this geometry supports two surface plasmon resonances, associated with polarisations aligned parallel and perpendicular to the symmetry axis of the nano-antenna. Diagonal ($D$) and anti-diagonal ($A$) polarisation states respectively excite a symmetric and an anti-symmetric mode (see Fig.~\ref{fig:fig1}-({\bf c})) -- analogous to the case of coupled harmonic oscillators. The optical retardation $\delta$ between these two optical modes can be adjusted to approximate $\pi$ by choosing the appropriate dimensions and periodicity of the nano-antennas, although these vary for different operating wavelengths~\cite{aieta:12}. The action of this metasurface, neglecting both absorption and reflection, in the Jones calculus (omitting a global phase delay) is given by:
\begin{align}
	\label{eq:jones_uniform}
	\sigma_z= \left(\begin{array}{cc}
	1 & 0 \\ 	
	0 & -1 \\ 
  \end{array}\right),
\end{align}
where $\exp{(i \pi)} = -1$ represents the relative optical retardation of $\pi$ between excited diagonal and anti-diagonal polarisation states. As can be seen, Eq.~(\ref{eq:jones_uniform}) represents the action of a half-wave plate in the A-D polarization basis. Let us now consider an array of such antennas that are rotated by an angle $\alpha$ about an axis orthogonal to the plate's surface. A straightforward Jones calculation determines the metasurface action as $R(-\alpha)\cdot\sigma_z\cdot R(\alpha)$, which in the \emph{circular polarisation} basis is given by
\begin{eqnarray}
	\label{eq:jones_general}
	{\cal M}_\alpha= \left(\begin{array}{cc}
    0 & e^{-2i\alpha}\\
    e^{2i\alpha} & 0 \\ 
  \end{array}\right),
\end{eqnarray}
where we assume $\ket{L}=\left(\begin{array}{c} 1\\ 0 \\  \end{array}\right)$ and $\ket{R}=\left(\begin{array}{c} 0\\ 1 \\  \end{array}\right)$ to be left-circular and right-circular polarisation states, respectively. Again a global phase ($\pi/2$) is omitted. As expected, in the circular polarisation basis, this proposed metasurface flips the polarisation state of an input beam from left ($L$) to right ($R$) circular polarisation or \emph{vice versa}, i.e. ${\cal M}_\alpha\cdot\ket{L}=\exp{(2i\alpha)}\,\ket{R}$. However, any rotation of the metasurface orientation does not change the output polarisation state $\ket{R}$, and the output polarisation state remains right circularly polarised (or left depending on the input handedness). Nevertheless, the rotated metasurface introduces a \emph{uniform} phase $\exp{(\pm2i\alpha)}$ that depends both on the input polarisation state and the orientation of the metasurface. This introduced phase, known as the Pancharatnam-Berry phase, is equal to half of the solid angle probed by the state evolution on the polarisation Poincar\'e sphere~\cite{pancharatnam:56,berry:87}. The sign of this phase is defined by the helicity of the input polarisation, positive for left-handed and negative for right-handed input polarised beams, corresponding to a clockwise or a counterclockwise path on the Poincar\'e sphere. However, in general, the orientation of the nano-antennas can be nonuniform, i.e. it can vary within the transverse plane. Here we consider the specific case where the orientation of the nano-antennas is dependent on the azimuthal angle of the polar coordinate system and the topological charge, i.e. $\alpha(\phi)=q\,\phi$.  In fact, the topological charge $q$ determines the amount of rotation of the nano-antennas for a full coordinate rotation ($\phi\rightarrow\phi+2\pi$). Different metasurfaces with topological charges of $q=1/2$, $1$, $3/2$, $10$ and $25/2$ are shown in Fig.~\ref{fig:fig1}-({\bf a}) and ({\bf b}). Apart from a small region very close to the origin, where the nano-antenna orientation is \emph{undefined} (a singularity), the metasurface introduces a nonuniform, helical staircase-like phase-front of $\exp{(\pm i\,2q\,\phi)}$ to the circulaly polarised ouput beam. Thus, the emerging beam possesses an OAM value of $\pm 2q$, with the sign depending on the input polarisation state. However, in practice the metasurface may not introduce a \emph{perfect} phase delay of $\pi$; thus its action on a circularly polarised basis is more complicated than discussed above. For such a plate, the action can be split into two parts; a portion of the beam that undergoes spin-to-orbital angular momentum conversion and gains an OAM value of $\pm2q$ and another portion that is not affected by the metasurface, remaining in the same polarisation and OAM states as the input beam. These two portions have orthogonal circular polarisation states, and thus they can be separated by means of a quarter-wave plate (QWP) followed by a polarising beam splitter (PBS).\newline

\begin{figure}[!h]
 \center
  \includegraphics[width=0.5\textwidth]{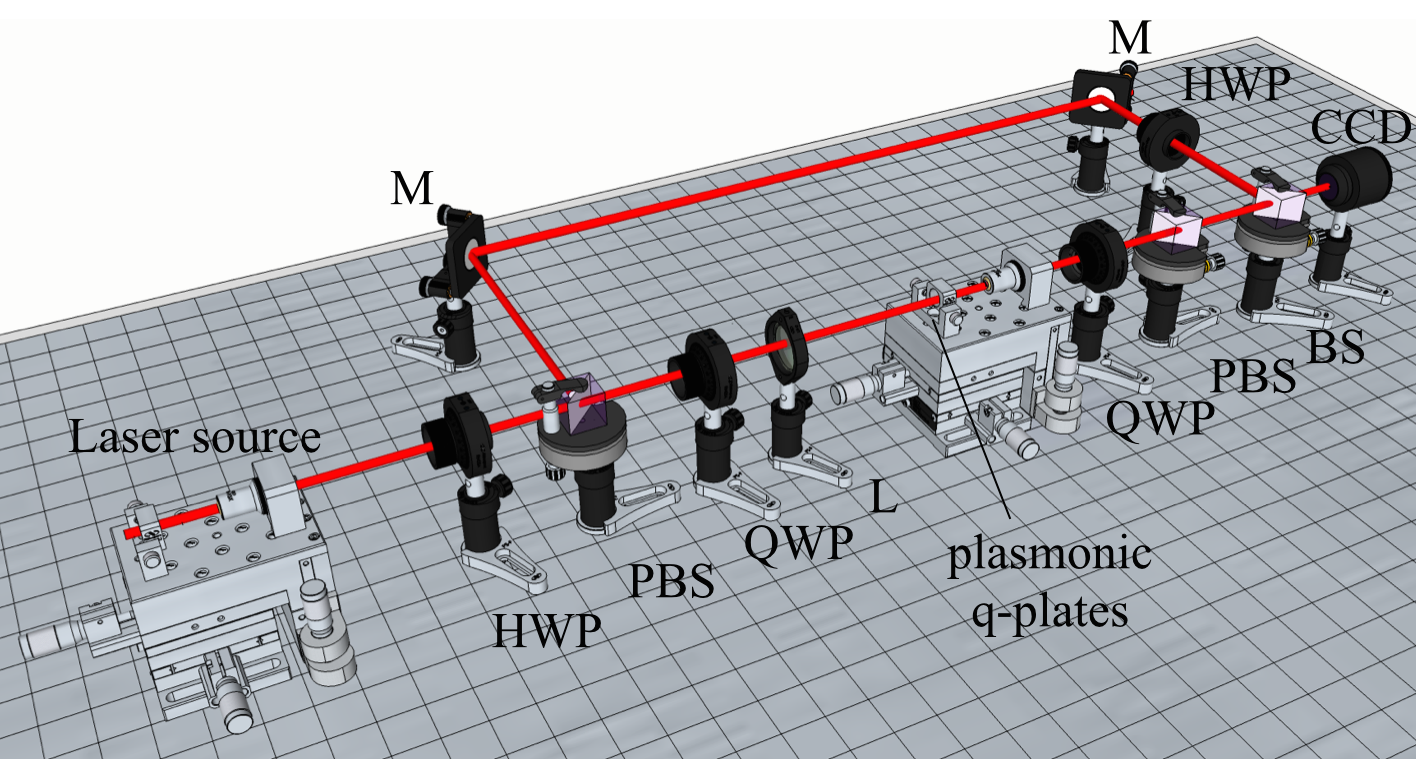}
  \caption{\label{fig:fig2} Experimental setup. M: mirror, L: lens, HWP: half-wave plate, QWP: quarter-wave plate, BS: beam splitter, PBS: polarizing beam splitter, CCD: CCD camera.}
\end{figure}
The gold metasurface was fabricated on a fused silica substrate through electron beam lithography in a positive bi-layer resist, followed by thermal metal evaporation and lift-off. Charge dissipation during the lithography step was achieved through a conductive layer consisting of 23~nm thick indium tin oxide (ITO). Typical dimensions for an individual nano-antenna are 142-150~nm arm width, 235-242~nm arm length and a thickness of 90-96~nm with a periodicity of 330~nm. This design provides an optical retardation of $1.35\pi$ radian between two linearly orthogonal polarisation states and a conversion efficiency of about 35\% at a wavelength of 780~nm. Here, we define the conversion efficiency as the ratio of the converted power to the total power of the transmitted beam, i.e. the sum of both the converted and the non-converted components transmitted out of the sample. Scanning electron microscope (SEM) images of several metasurface plates having different topological charges are shown in Fig.~\ref{fig:fig1}-({\bf b}). As can be seen in Fig.~\ref{fig:fig1}-({\bf c}), the shape of nano-antenna does not form a perfect L-shape. Because the individual antennas are not perfectly L-shaped, their birefringence between the resonances of the $A$ and $D$ polarizations is relaxed somewhat, which affects the metasurface's conversion efficiency but does not compromise the working principle of the device. Similarly, the small number of nano-antennas missing from the device surface affects only the intensity profile of the emerging beam.
\begin{figure*}[!ht]
 \center
  \includegraphics[width=0.8\textwidth]{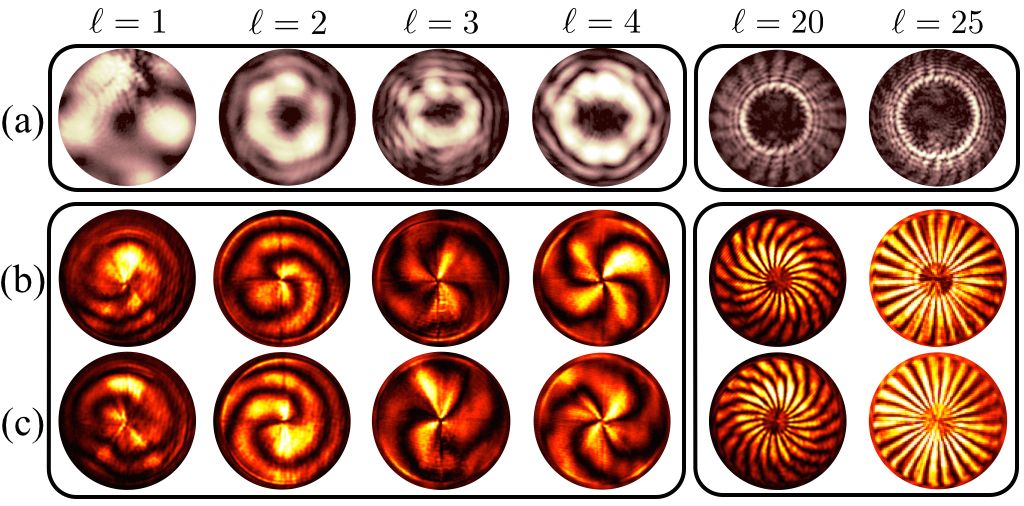}
  \caption{\label{fig:fig3} Intensity profile and recorded interference patterns of the beam generated by the plasmonic metasurfaces for different OAM values. (a) The intensity profile of the converted component of the transmitted light. The central null field at the origin results from the phase singularity. (b)-(c) The interference pattern of the generated beams with a spherical wave. The (b) and (c) rows correspond to left and right-circularly polarised input beams. The number of intertwined helices reveals that the transmitted beams possess OAM of $\ell=\pm1$, $\pm2$, $\pm3$, $\pm4$, $\pm20$ and $\pm25$.}
\end{figure*}
%

The experimental setup shown in Fig.~\ref{fig:fig2} is used to record both the intensity and the wavefront of the beams generated by the metasurfaces. Measurements are carried out using a tunable diode laser, working in the wavelength range of 760~nm to 790~nm. The single-mode laser beam is spatially cleaned through coupling into a single mode optical fibre. The beam is circularly polarised by means of a PBS followed by a rotated QWP. The reflected beam from the PBS, which is vertically polarised, is used as a reference beam in the Mach-Zehnder interferometer, to record the wavefront of the generated beam. In order to record the interference pattern, the polarisation of the reference beam is rotated to horizontal polaristion by means of a half-wave plate. On the sample arm, a convex lens with a focal length of $200$~mm is used to reduce the beam size to be comparable with the diameter of the \emph{patterned} metasurface ($\simeq100~\mu$m). A $10$X microscope objective is used to collimate the transmitted beam. The transmitted beam contains both the converted and non-converted parts, with orthogonal circular polarisations, e.g. for a left handed circularly polarised input beam the converted and non-converted beams have right- and left-circular polarisations, respectively.  A sequence of a rotated QWP at an angle of $\pi/4$ and a PBS is used to separate the converted and non-converted light, for simultaneous power measurements.
The intensity profile of the converted beam, which possesses OAM, is then captured by a (DCU223C) CCD camera. Figure~\ref{fig:fig3}-({\bf a}) shows the intensity profile of the converted beams generated by the plasmonic metasurfaces shown in Fig.~\ref{fig:fig1}-({\bf b}). As can be seen, the emerging beams possess a doughnut shape with a \emph{null} intensity at the origin. This null intensity, which is caused by the phase singularity at the beam centre, is characteristic for beams with a helical phase front. The plasmonic plate introduces such a helical phase-front of $\exp{(\pm i\,2q\,\phi)}$ into the Gaussian input beam. At the metasurface exit face, this beam is proportional to $\exp{(-r^2/w_0^2)}\,\exp{(\pm i\,2q\,\phi)}$, where $r$ is the radial distance in polar coordinates and $w_0$ is the beam waist of the input Gaussian beam. This is a subfamily of Hypergeometric-Gaussian beams that reduce to a  superposition of two Bessel-Gaussian beams~\cite{karimi:07,sacks:98} and is not shape-invariant under free-space propagation, as explained in Ref.~\cite{karimi:07}. The OAM value of the emerging beam is determined by recording the interference pattern with a reference beam. The interference of a spherical reference wave and a beam carrying OAM value of $\ell$ results in a helical pattern with $\ell$ intertwined lobes. Switching the input polarisation state from left-circular to right-circular polarisation does not change the intensity profile, but changes the sign of the OAM value $\ell$. Thus, it reverses the twisting direction of the intertwined helices of the interference pattern, without affecting the number of helices. Figure~\ref{fig:fig3}-({\bf b}) and ({\bf c}) show interference patterns of the converted beams from the plasmonic metasurfaces (shown in Fig.~\ref{fig:fig1}-({\bf b})) for both left-circular and right-circular input polarisation states. These confirm that the output beams possess OAM values equal to twice the metasurface topological charge, i.e. $|\ell|=2q$ as expected from our previous discussion.

Finally, the conversion efficiency of the device is measured for the wavelength range from 760~nm to 790~nm. In our definition, both reflections and absorptions by different optical components, including the metasurface, are not considered. This distinguishes the \emph{purity} of the generated beam from absorption and reflections off optical components in the experimental set-up, as this \emph{purity} is generally the limiting factor. Conversion efficiencies of $(8.6\pm0.4)$\% at a wavelength of 780~nm are measured for different metasurface plates. This conversion efficiency is reasonably spectrally broad, for example the $q=2$ metasurface showed a change of only 0.4\% across the 30 nm wavelength range. \newline

In summary, we designed and fabricated plasmonic metasurfaces with arbitrary integer and half-integer topological charges capable of generating beams of light carrying integer magnitudes of OAM. The largest OAM value tested was 25, but there is no fundamental upper bound on the $\ell$ magnitudes that could be achieved by these plasmonic metasurfaces. We experimentally investigated spin to OAM conversion at various $\ell$ values by recording the intensity and phase profiles of the beams generated by these metasurfaces. About 9\% of the transmitted light was converted. This technology would introduce ultra-thin OAM generators into integrated devices, which could have applications for nano-scale sensing~\cite{hell:07}, classical~\cite{gibson:04,bozinovic:13} and quantum communications~\cite{vallone:14,mirhosseini:14}. 

The authors thank Hammam Qassim for fruitful discussions, and acknowledge the support of the Canada Excellence Research Chairs (CERC) Program.

\end{document}